\documentclass{article}
\usepackage{spconf,amsmath,graphicx}
\usepackage{doi}

\title{ACCELERATED FBP FOR COMPUTED TOMOGRAPHY IMAGE RECONSTRUCTION}

\name{Anastasiya Dolmatova \supok{1,2,}\sthanks{This work was supported by the Russian Foundation for Basic Research (18-29-26020, 18-29-26033).}, Marina Chukalina\supok{1,2,3} and Dmitry Nikolaev\supok{1,2}}

\address{ 
	\supok{1}IITP RAS (Kharkevich Institute), Moscow, Russia;\\
	\supok{2}Smart Engines Service LLC, Moscow, Russia; \\
	\supok{3}FSRC "Crystallography and Photonics", Moscow, Russia.}

\begin{document}

$\copyright$ 2020 IEEE. Personal use of this material is permitted. Permission from IEEE must be obtained for all other uses, in any current or future media, including reprinting/republishing this material for advertising or promotional purposes, creating new collective works, for resale or redistribution to servers or lists, or reuse of any copyrighted component of this work in other works.
	%
	\maketitle
	\begin{abstract}
		Filtered back projection (FBP) is a commonly used technique in tomographic image reconstruction demonstrating acceptable quality. The classical direct implementations of this algorithm require the execution of $\Theta(N^3)$ operations, where $N$ is the linear size of the 2D slice. Recent approaches including reconstruction via the Fourier slice theorem  require $\Theta(N^2\log N)$ multiplication operations. In this paper, we propose a novel approach that reduces the computational complexity of the algorithm to $\Theta(N^2\log N)$ addition operations avoiding Fourier space. For speeding up the convolution, ramp filter is approximated by a pair of causal and anticausal recursive filters, also known as Infinite Impulse Response filters. The back projection is performed with the fast discrete Hough transform. Experimental results on simulated data demonstrate the efficiency of the proposed approach. 
	\end{abstract}

	\section{Introduction}
	\label{sec:intro}
	
	X-ray computed tomography (CT) is a highly-regarded technique for medical diagnostics~\cite{rubin2014computed, smelkina2017computed}, industrial quality control~\cite{carmignato2018industrial, wang2015industrial}, material science research~\cite{salvo20103d, baruchel2000x} etc. The rapid increase of the CT scanner resolution necessitates processing a huge amount of data, so performance of the classical reconstruction method does not meet current industry demands~\cite{simonov2017non}.  Filtered Back Projection (FBP),  is a commonly utilized analytic image reconstruction algorithm that has  computational complexity ($\Theta(N^3)$, where $N$ is the linear size of the 2D slice).
	The most popular approach to make FBP algorithm faster is based on the projection-slice theorem (\textit{or} central slice theorem)~\cite{potts2000new}. Several algorithms use the fast Fourier transform on an inhomogeneous grid, followed by the transition from polar coordinates to Cartesian coordinates in Fourier space using interpolation. The quality of reconstruction in this case strongly depends on the choice of interpolation method. A detailed analysis of arising artifacts that distort the output image is given in~\cite {Potts2002}. In~ \cite{andersson2005fast} the invariance properties of the Radon transform and its dual has been used to construct a method of inversion based on log-polar representations.
	
	A completely different approach  was reported in~\cite{basu2000n, xiao2002n}. The main idea  is to reconstruct not the whole image, but use the properties of Radon transform to calculate sinograms corresponding to the four quadrants of the image, and  reconstruct them individually. In this case, according to the Nyquist--Shannon theorem, only half projections are required to reconstruct a quadrant of the image without losing the quality. Splitting of the image into quadrants can be continued sequentially until the size of the independently reconstructed sections reaches a value of 1 pixel. In this case, the algorithm requires $\Theta(N^2\log N)$ multiplications. The shortcoming of this method is a large number of intermediate interpolations, which can lead to the accumulation of the error ~\cite{basu2001error}.
	
	All described methods require $\Theta(N^2\log N) $ multiplication operations. In this paper, we propose a novel approach, which allows reconstructing the image in $\Theta(N^2\log N)$ addition operations and $\Theta(N^2)$ multiplication operations.
	
\section{Filtered Back Projection}
\label{sec:fbp}
\begin{figure}[htb]
		\begin{minipage}[b]{.48\linewidth}
			\centering
			\centerline{\includegraphics[width=4.0cm]{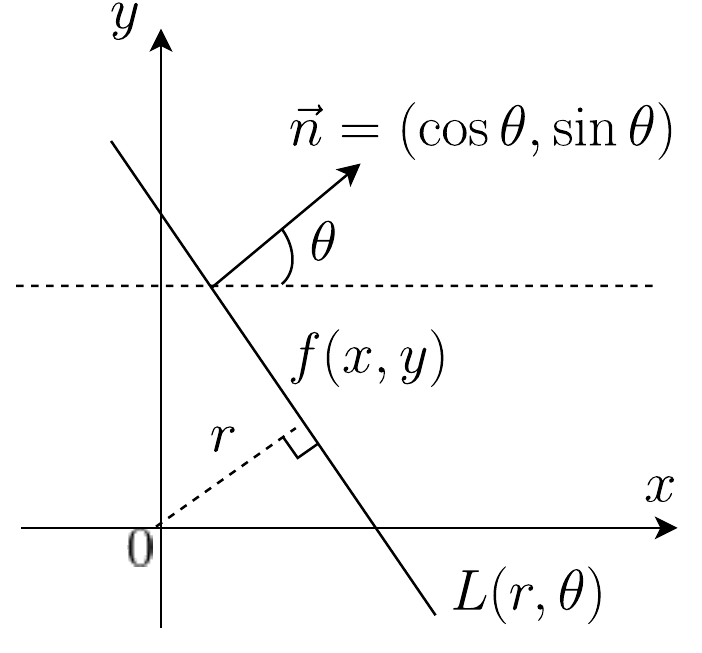}}
			\centerline{(a) $(r,\theta)$-parameterization}\medskip
		\end{minipage}
		\hfill
		\begin{minipage}[b]{0.48\linewidth}
			\centering
			\centerline{\includegraphics[width=4.0cm]{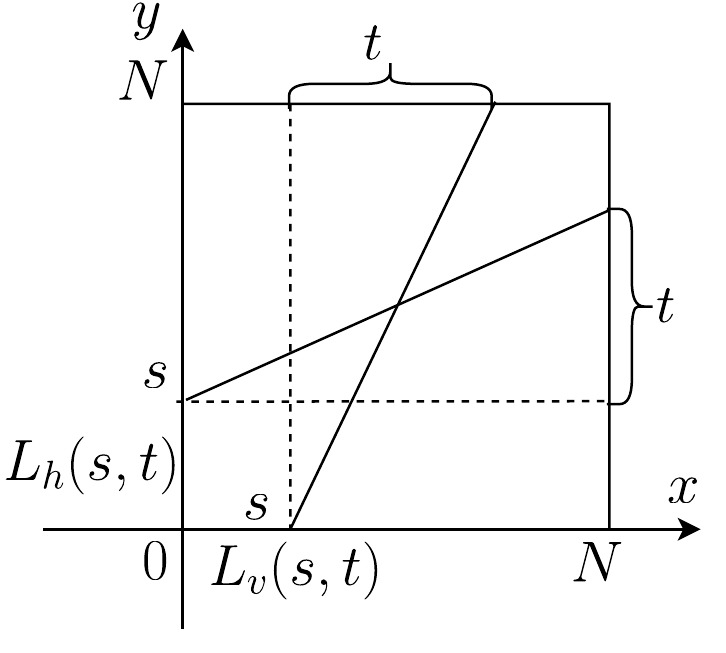}}
			\centerline{(b) $(s,t)$-parameterization}\medskip
		\end{minipage}
		\caption{Different parameterizations of straight lines on the plane: (a)  slope of the normal $\theta$ and the distance to the origin $r$; (b) x- or y-intersect $s$ and horizontal or vertical shift $t$ (for mostly vertical lines $L_v$ and mostly horizontal lines $L_h$, respectively).}
		\label{fig1}
	\end{figure}

	Let us briefly recall the basis of the FBP method. The Radon transform defined on the space of straight lines $L$ is the integral transform 
	\begin{equation}
		\mathcal{R}[f](L)=\int_{L}f(x,y)dl,
		\label{eq1}
	\end{equation}
	where $f(x,y)$ is some finite continuous function defined on the plane.
	In computed tomography, a straight line is usually given by the slope of the normal $\theta$ and the distance from the origin $r$ (see Figure~\ref{fig1}a). The projection of the function $f(x,y)$ is the set of all points in its Radon transform corresponding to a certain angle $\theta$:  
	\begin{equation}
		{p}_{\theta }(r)\equiv {{\left. \mathcal{R}[f](\alpha ,r) \right|}_{\alpha =\theta }}.
		\label{eq2}
	\end{equation}
	
	The essence of the FBP method is the sequential application of two operations. At the first step, the convolution of the projections with ram-lak (\textit{or} ramp)~\cite{ramachandran1971three} filtering function is performed:
	\begin{equation}
		\widetilde{p}_{\theta}(r)=p_{\theta}(r) * h(r),
		\label{eq3}
	\end{equation}
	where 
	\begin{equation}
		h(r)=\int_{-\infty}^{\infty}|\omega|e^{2\pi i\omega r}d\omega.
		\label{eq4}
	\end{equation}
	Then, conventional back projection is applied:
	\begin{equation}
		\widetilde{f}(x,y)=\mathcal{B}[\widetilde{p}](x,y)=\int_{0}^{2\pi }\widetilde{p}_\theta(r)d\theta,
		\label{eq5}
	\end{equation}
	where $r=x\cos\theta+y\sin\theta$. 
	
	In real measuring systems, it is not possible to obtain a continuous set of projections; therefore, all integral operators are replaced by appropriate summation, and the continuous convolution is converted into one-dimensional linear filtering. In computed tomography, the number of projection angles $P$ is chosen to be of the same magnitude as the image size $N$. In this case,  convolution can be performed with $\Theta(N^3)$ operations in the image space or in $\Theta(N^2\log N) $ operations using fast Fourier transform (FFT). The back projection requires $\Theta(N^3)$ operations. 
	
\section{Fast recursive filtering }
\label{sec:filtering}
	
	Consider an ideal ramp filter with an impulse response given by the expression~(\ref{eq4}). Generally, it has a singularity at the point $r=0$~\cite{zeng2014revisit}. However, in real cases, the spectrum of measured projections is  band-limited with the bandwidth $2W$ ($|\omega|<W $).
	For a discrete signal, $W = 0.5/\Delta$, where $\Delta$ is the sampling rate. Without loss of generality,  we can set $\Delta = 1$ by choosing the appropriate coordinates system. In this case, the impulse response of the discrete filter~(\ref{eq4} takes the form
	\begin{equation}
		{h}(n)=2\int\limits_{0}^{1/2}|\omega|e^{2\pi i\omega n}d\omega=\frac{sinc(\pi n)}{2}-\frac{sinc^2(\pi n/2)}{4}
	\label{eq6}
	\end{equation}
	where  $sinc(r)=\sin(r)/r$.
	Simplifying of the latter expression yields 
	\begin{equation}
		{h}(n)=
		\begin{cases}
			1/4, & n=0,\\
			0, & n \text{ even},\\
			-1/(n\pi)^2, & n \text{ odd}.\\
		\end{cases}
		\label{eq7}
	\end{equation}
	
	The discrete convolution of the projection $p_\theta$ with kernel~(\ref{eq7}) can be recast as a finite impulse response filter (\textit{or} FIR filter):
	\begin{equation}
		\widetilde{p}_\theta(n)= p_\theta \ast h=\sum_{k=n-L_0}^{k=n+L_0}p_\theta(n-k)h(k),
		\label{eq8}
	\end{equation}
	where $L=2L_0+1$ is the length of the filter kernel.
	
	Even though the function $h(n)$ decays quadratically with $n$, a decrease in the length of the filter kernel leads to a significant distortion of the reconstructed image. To achieve the minimum error, the kernel length should be the same order of magnitude as $N$. Thus, although formally calculating the convolution~(\ref{eq8}) requires $\Theta(N^2)$ operations, in the real cases, the factor for $N^2$ is very large. The issue of reducing the computational complexity of the convolution is essential for many signal processing problems.
	
	In paper~\cite{deriche1987using}, the authors presented a group of computationally effective methods for approximating a Gaussian filter, its first and second derivatives using filters with infinite impulse response (IIR filters). The difference expression describing the discrete IIR filter has the form:
	\begin{equation}
		\widetilde{p}_\theta(n) = \sum_{k=0}^{M-1}b_k p_\theta(n-k) - \sum_{k=1}^{Q}a_k {\widetilde{p}_\theta}(n-k),
		\label{eq9}
	\end{equation}
	where $M$ and $Q$ are the feedforward and feedback filter orders, $a_k$ and $b_k$ are the coefficients characterizing the filter. The advantage of the IIR filter is that for $M\ll N$ and $Q\ll N$, $\Theta(N^2)$ operations are required to process an image of size $N\times N $, which is significantly less than for an FIR filter. 	One can note that the impulse response of the filter~(\ref{eq7}) is unidirectional, while the impulse response of the ramp filter is symmetric ($r(n) = r(-n)$). A symmetric recursive filter can be represented as the sum of the casual and anticasual component~\cite{deriche1993}. 
	
	The IIR filter is constructed so that its impulse response is equal to the impulse response of the FIR filter $h(n)$. Then, the impulse response~(\ref{eq7}) can be rewritten as the sum of casual and anti-casual components:
	\begin{equation}
		{h}^+=
		\begin{cases}
			h(0)/2, &n=0,\\
			h(n), &n>0,\\
			0, &n<0.\\
		\end{cases}	
		{h}^-=
		\begin{cases}
			h(0)/2, &n=0,\\
			h(n), &n<0,\\
			0, &n>0.\\
		\end{cases}
		\label{eq10}
	\end{equation}
	Thus, 
	\begin{equation}
		\widetilde{p_\theta}=p_\theta \ast h^+ + p_\theta \ast h^-.
		\label{eq11}
	\end{equation}
	The resulting function $\widetilde{p}_\theta$ can be presented as the sum of the outputs of two recursive filters:
	\begin{equation}
		\widetilde{p}_\theta(n) = \widetilde{p}^+_\theta(n) + \widetilde{p}^-_\theta(n),
		\label{eq12}
	\end{equation}
	where
	\begin{equation}
		\widetilde{p}^\pm_\theta(n) = \sum_{k=0}^{M-1}b_k^\pm p_\theta(n\mp k) - \sum_{k=1}^{Q}a_k^\pm \widetilde{p}_\theta(n\mp k).
		\label{eq13}
	\end{equation}
	
%
%
%
%

	Due to the symmetry, the coefficients $a_k^+ = a_k^-$ and $b_k^+ = b_k^-$, so it is enough to determine the coefficients only for a casual filter. Coefficients can be found by  minimizing the mean-square error between the impulse response of the FIR filter~(\ref{eq8}) and the IIR filter~(\ref{eq9}). One can use any optimization algorithm, for instance,  the Powell conjugate gradient method~\cite{powell1964efficient} or the simplex method~\cite{gao2012implementing}.
	It is worth to note that proposed scheme could be applied not only to the ramp filter, but to any other filter used in FBP approach (e.g. \cite{horbelt2002discretization}).

	
\section{Fast back projection}
\label{sec:fastfbp}

\subsection{$(s, t)$-parametrization}
\label{ssec:st}
	Introduce $(s, t)$-parameterization of the line so that some point $(x_0,y_0)$ on the original image plane $(x,y)$ defines a line on the parameter plane $(s,t)$. The set of projections in $(s,t)$-space is sometimes called the linogram~\cite{edholm1987linograms}. 
	
	Let the function $f(x, y)$ be given in the squared area $0\le x \le N$, $0 \le y \le N$. We divide the set of all lines into four classes: mostly vertical lines with positive shift $L_v^+$ ($3\pi/4\le\theta<\pi$); mostly vertical lines with negative shift $L_v^-$ ($0\le\theta<\pi/4$); mostly horizontal lines with positive shift $L_h^+$ ($\pi/2\le\theta<3\pi/4$); mostly horizontal lines with negative shift $L_h^-$\\ ($\pi/4\le\theta<\pi/2$).

	Parameters $s$ and $t$ specify the coordinates of two points of the line lying on the vertical (for $L_h^\pm$) or horizontal (for $L_v^\pm$) boundaries (see Figure~\ref{fig1}b). Parameter $t$ takes values from $-N$ to $0$ for $L_h^-$ and $L_v^-$; and from $0$ to $N$ for $L_h^+$ and $L_v^+$. Thus, the final linogram contains four $N \times N$ images for all types of lines. 

\subsection{$(r, \theta)$--$(s, t)$ transition}
\label{ssec:rt-st}
	There is one-to-one relationship between linogram coordinates $(s,t)$ and original sinogram coordinates $(r,\theta)$:
	\begin{equation}
		\tan\theta=-\left(N/t\right)^{p}, \qquad r=sN/\sqrt{t^2+N^2},
		\label{eq14}
	\end{equation}
	where $p=1$ for $L_h^\pm$ and $p=-1$ for $L_v^\pm$. Transition from sinogram to linogram using  linear interpolation requires  $\Theta(N^2)$ operations.
	
	However, a careless transition from $(r, \theta)$ to $(s,t)$ variables can lead to the appearance of an error related to the violation of the rotational invariance of Radon transform in $(s,t)$-space, obtained for the squared image domain.
	Let us consider the projection in the linogram for $L_h^+$ with shift $t$. In the sinogram, this projection correspond to the projection with the angle of inclination $\phi_t = \theta_t - \pi/2 = \arctan(t/N)$ (see Eq.~(\ref{eq14})). The length of the corresponding line is $N/\cos\phi$. One can note, that this length is not constant and depends on the angle $\phi$. Since Radon transform in squared domain should preserve the Radon invariant (the sum of the values in any row is equal to the total sum in the image), the projection amplitude is underestimated relative to the conventional Radon transform (which is equal to the sinogram obtained by CT scanner) by a factor of $k_t=1/\cos\phi_t$. Thus, each linogram projection  is "stretched" relative to the corresponding sinogram projection by the same factor $k_i$.
	Expressing the scaling coefficient explicitly yields
	\begin{equation}
		k_t = \sqrt{1+t^2/N^2}.
		\label{eq15}
	\end{equation}
	It is important to keep this parameter in mind when converting linogram to sinogram.
	
	\subsection{Back projection in $(s, t)$-coordinates}
	\label{ssec:backprojection}
	
	An important feature of $(s,t)$-parameterization is morphological symmetry:  each point in the original image  corresponds to a straight line in the linogram, and each point in linogram  corresponds to a straight line in the image. Such symmetry allows us to establish a connection between the forward projection operator (Radon transform) $\mathcal{R}[f](s, t)$ and the corresponding back projection operator $\mathcal{B}[p](x, y)$. 
	Denote parts of linogram corresponding to the introduced classes of lines as $P_h^+(s,t)$, $P_h^-(s,t)$, $P_v^+(s,t)$ and $P_v^-(s,t)$. One can note, that forward projection operators for mostly horizontal lines $\mathcal{R}_h^\pm[f]$ can be obtained from the corresponding operators for mostly vertical lines $\mathcal{R}_v^\pm[f]$ by preliminary transposing the image:
	\begin{equation}
		P_h^\pm(s,t) = \mathcal{R}_h^\pm[f](s,t) = \mathcal{R}_v^\pm[f^T](s,t).
		\label{eq16}
	\end{equation}
	Rewriting the expressions for the forward~(\ref{eq1}) and back~(\ref{eq5}) projection operators in $(s,t)$-coordinates yields
	\begin{align}
		P_v^\pm(s,t)&=\mathcal{R}_v^\pm[\widetilde{f}](s,t)=\int_{0}^{N}\widetilde{f}\left(s+\frac{t}{N}y,y\right)dy,
		\label{eq17}
		\\
		\widetilde{f}_v^\pm(x,y)&=\mathcal{B}_v^\pm[P_v^\pm](x,y)=\pm\int_{0}^{N}P_v^\pm\left(y-\frac{x}{N}t,t\right)dt.
		\label{eq18}
	\end{align}
	
	Comparing the expressions~(\ref{eq17}) and~(\ref{eq18}), one can note that the operator $\mathcal{B}_v^\pm$ defining a mapping from  $(s,t)$-space to $(x, y)$-space can be expressed via the forward projection operator:
	\begin{equation}
		\mathcal{B}_v^\pm[P_v^\pm](x,y) = \pm\mathcal{R}_v^\pm[P_v^\pm](-x,y).
		\label{eq19}
	\end{equation}
	
	Thus, the back projection operation in $(s,t)$-space is equivalent to the forward projection operation with the change of sign of one of the parameters. Similarly,  operator $\mathcal{B}_h^\pm$ can be expressed as
	\begin{equation}
		\mathcal{B}_h^\pm[P_h^\pm](x,y) = \pm\left\{\mathcal{R}_h^\pm[P_h^\pm](-x,y)\right\}^T.
		\label{eq20}
	\end{equation}

\subsection{Discrete space. Fast Hough transform}
\label{ssec:fht}

	    \begin{figure}[tb]
		\begin{minipage}[b]{.48\linewidth}
			\centering
			\centerline{\includegraphics[width=4.5cm]{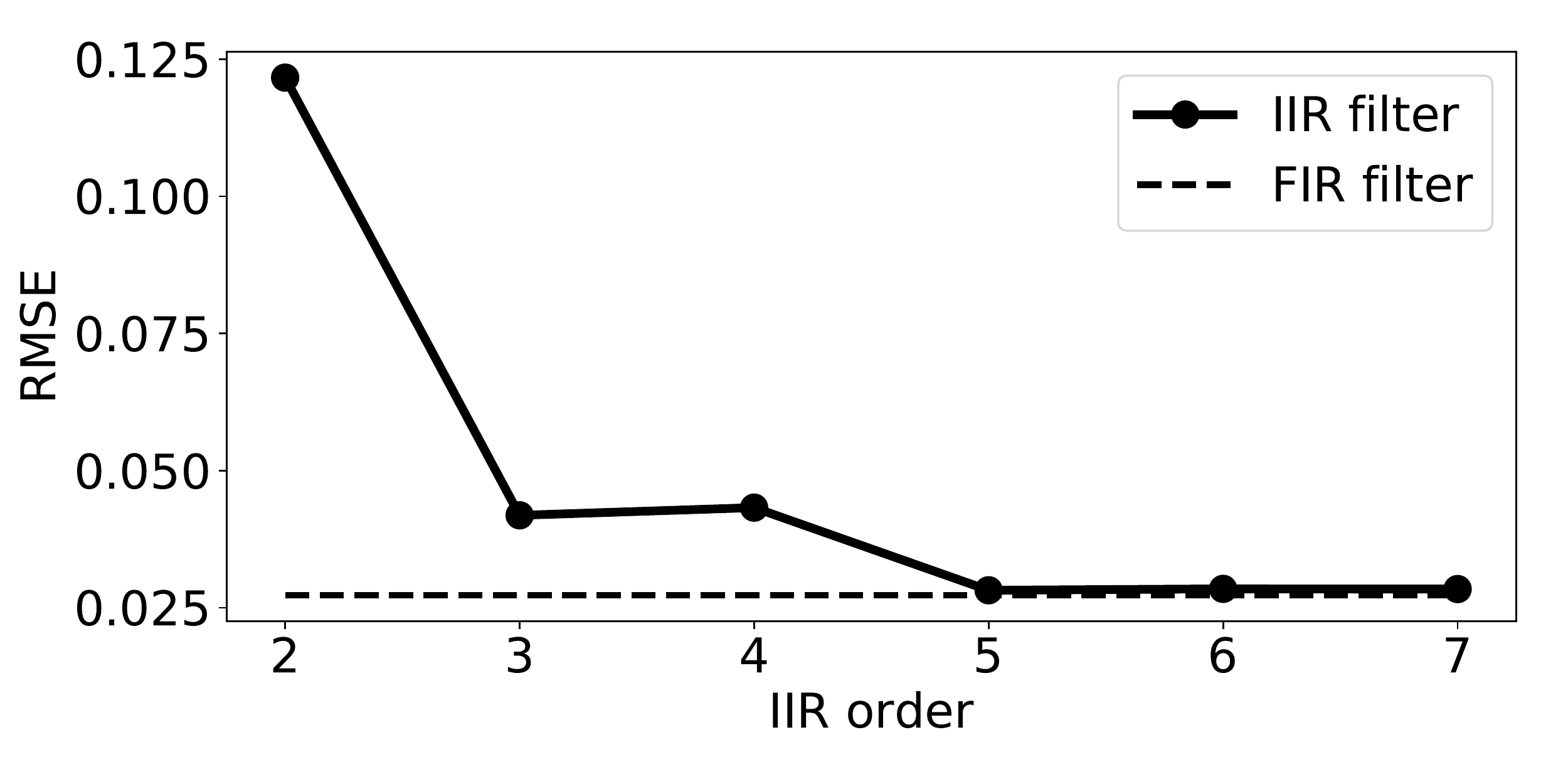}}
			\centerline{(a)}
		\end{minipage}
		\hfill
		\begin{minipage}[b]{0.48\linewidth}
			\centering
			\centerline{\includegraphics[width=4.5cm]{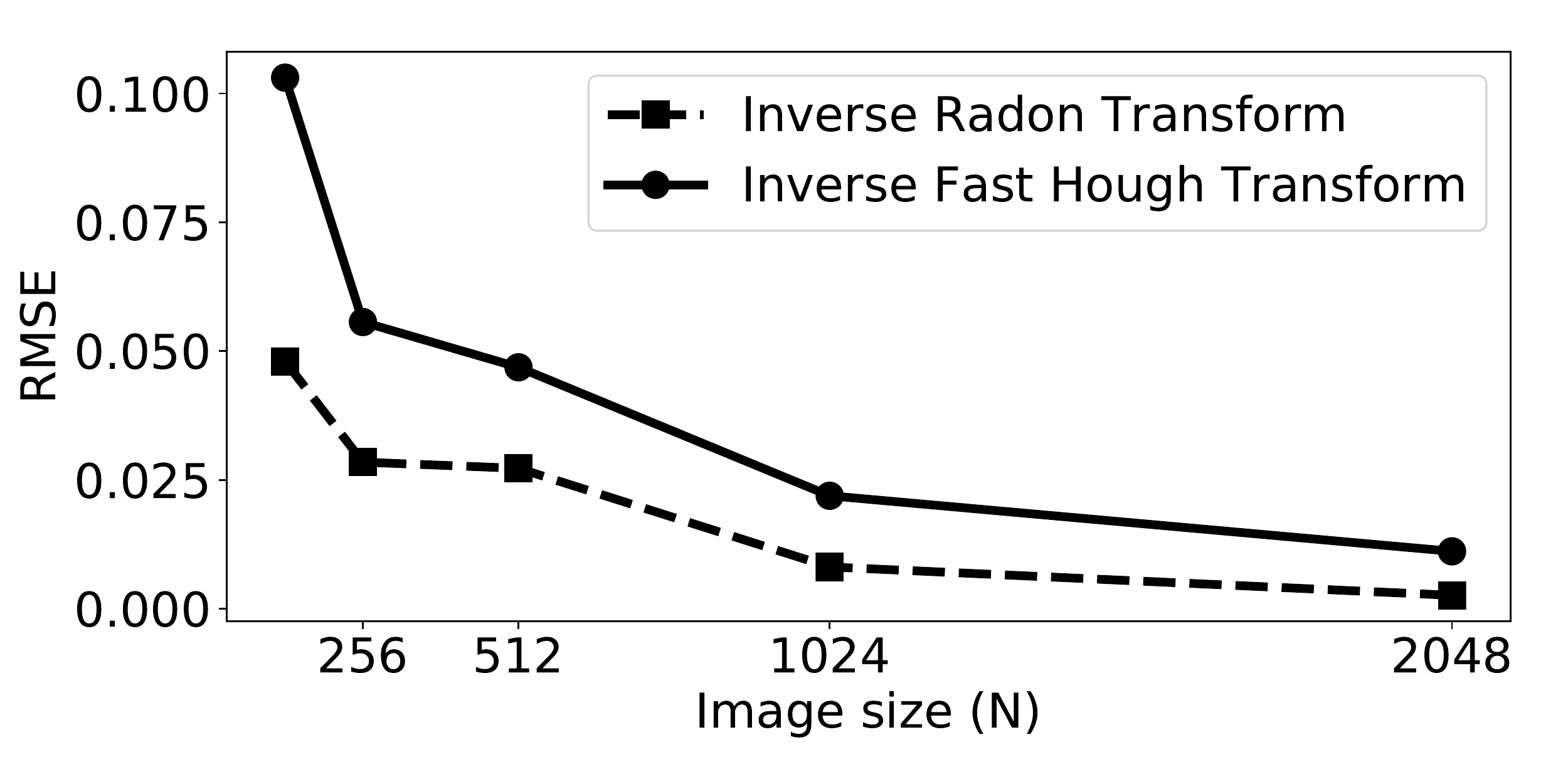}}
			\centerline{(b)}
		\end{minipage}
		\caption{Difference in the RMSE: (a)  IIR order dependence; (b) image size dependence (FIR filter).}
		\label{fig2}
	\end{figure}
	In a discrete space, the Radon transform of a function along a given line can be approximated by the sum of the functions values at points belonging to the discrete approximation of this line. With the appropriate choice of approximation, the  time required for the calculation of the discrete projection operator can be significantly reduced. In the work~\cite {brady1998fast}, M.~Brady noted that discrete representations of two lines with close slopes have a significant number of common points. In this case, there is no need to calculate the repeating section twice to find the sum along each of these lines. Brady proposed sequentially calculating partial sums for segments of length $2^i, i = 1 ... \log_2(N + 1)$. Paper~\cite{ershov2018generalization} presents  a recursive implementation of the described algorithm for the lines approximated by so-called dyadic patterns. This algorithm is also known as Fast Hough Transform (FHT). 
    In this implementation, the results are obtained separately for each type of lines ($L_h^\pm, L_v^\pm$). 
	The computational complexity of the algorithm is  $\Theta(N^2\log N)$ operations. Moreover, all these operations are summation, not multiplication.
	
	An asymptotically fast SART algorithm  based on an implementation of the fast Hough transform was proposed in~\cite{prun2013computationally}.
	\subsection{Brady approach for back projection (Inverse Fast Hough Transform)}
\label{ssec:fiht}

	According to the expressions~(\ref{eq19}) and~(\ref{eq20}), the back projection operator can be presented as a forward projection operator with the change of sign of one of the parameters. Consequently, one can apply the Brady approach:

	\begin{align}
		\begin{split}
			\widetilde{f}_h^\pm(x,y) &= \mathcal{R}[P_h^\pm(t,s)](-x,y),\\
			\widetilde{f}_v^\pm(x,y) &= \left\{\mathcal{R}[P_v^\pm(t,s)](-x,y)\right\}^T,
		\end{split}
		\label{eq21}
	\end{align}
	The final reconstruction is the sum of four images:
	\begin{align}
		\widetilde{f}(x,y) = \widetilde{f}_h^+(x,y)+\widetilde{f}_h^-(x,y)  + \widetilde{f}_v^+(x,y)  + \widetilde{f}_v^-(x,y).
		\label{eq22}
	\end{align}
	

\section{Results}
\label{sec:results}

	Experiments were conducted on Shepp-Logan phantom. We reconstructed the images with FIR and IIR filters ($Q=M$, $N=512$, $P=512$). IIR filter coefficients were found with the Simplex method. We investigated two parts of the proposed algorithm one by one. Root-Mean-Square-Error (RMSE) dependence on IIR filter order is presented in Fig. 2a. Radon transform is used for back projection. Dependence between RMSE and image size for Radon and Fast Hough back projection is presented in Fig. 2b. FIR filter is used for both (Radon and Hough) cases.
	
	\begin{figure}[tb]
	\begin{minipage}[b]{.48\linewidth}
		\centering
		\centerline{\includegraphics[width=2.5cm]{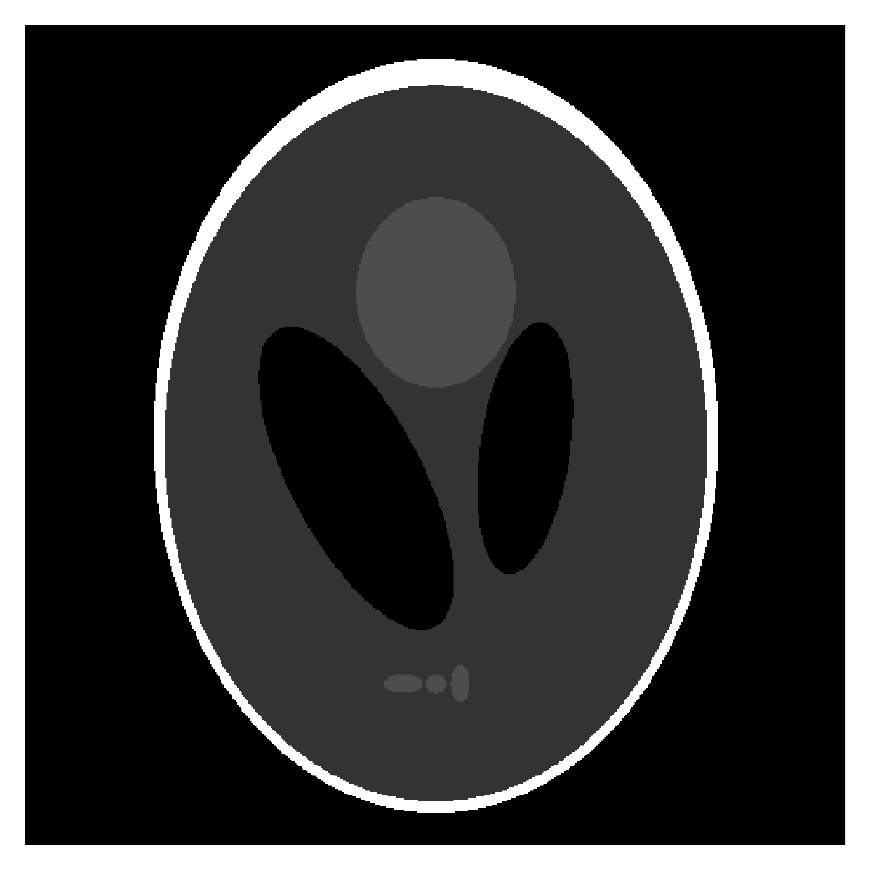}}
		\centerline{(a)}\medskip
	\end{minipage}
	\hfill
	\begin{minipage}[b]{0.48\linewidth}
		\centering
		\centerline{\includegraphics[width=2.5cm]{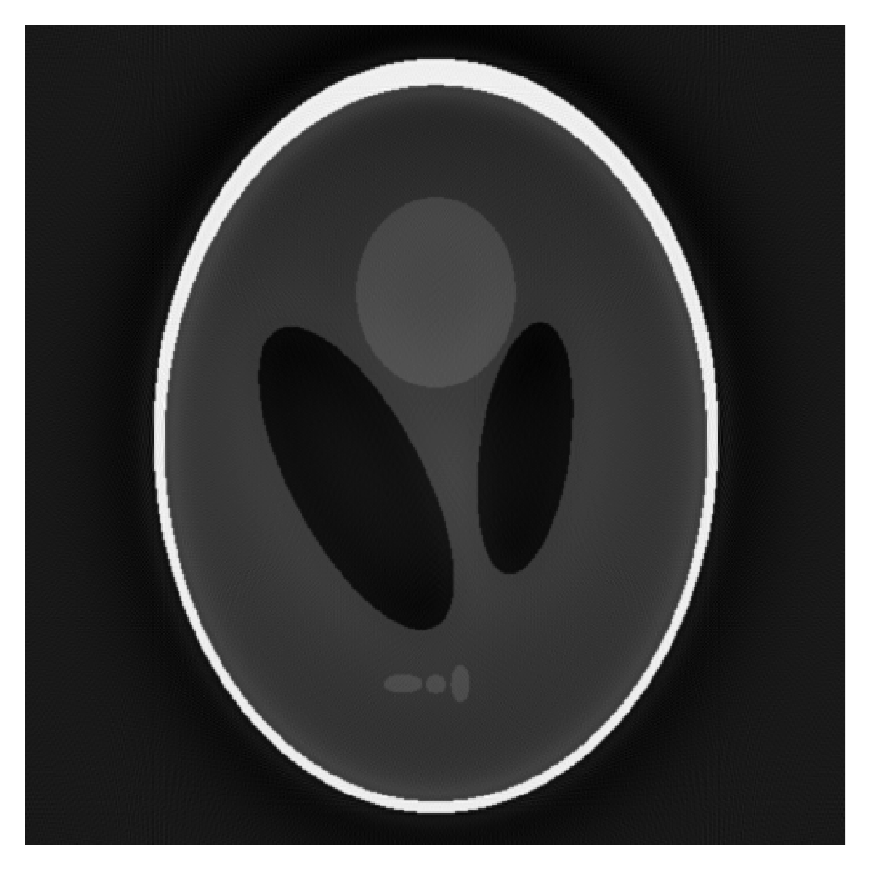}}
		\centerline{(b)}\medskip
	\end{minipage}
	\begin{minipage}[b]{.48\linewidth}
		\centering
		\centerline{\includegraphics[width=2.5cm]{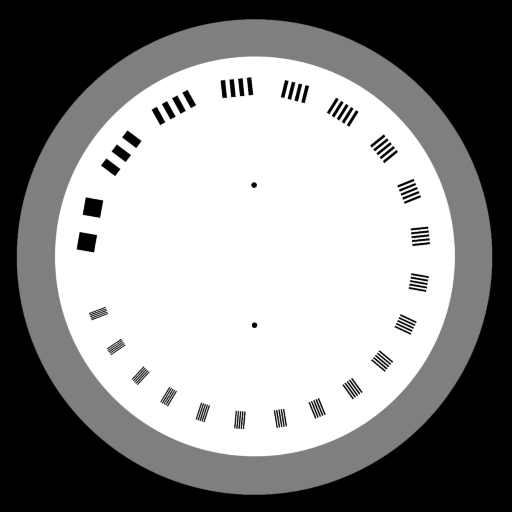}}
		\centerline{(c)}\medskip
	\end{minipage}
	\hfill
	\begin{minipage}[b]{0.48\linewidth}
		\centering
		\centerline{\includegraphics[width=2.5cm]{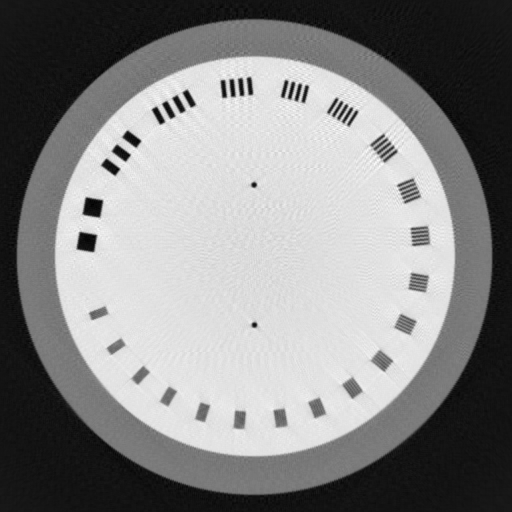}}
		\centerline{(d)}\medskip
	\end{minipage}
	\caption{Shepp-Logan (a) and CATPHAN (c) phantoms and the reconstructed images (b, d) obtained with the proposed algorithm, correspondingly.}
	\label{fig3}
\end{figure}
	
      Reconstructed image using the proposed algorithm ($N=1024$, $P=1024$, $Q=M=3$) is presented in Figure~\ref{fig3}. The algorithm requires $\Theta(N^2)$ operations for interpolation,  filtering and formation of the output images and $\Theta(N^2\log N)$ summations for back projection.

\section{Conclusion}
\label{sec:concl}
	In this paper we propose a novel fast algorithm to reconstruct the image from tomographic projections. Following FBP strategy we apply IIR filter with precalculated coefficients to the input sinogram to speed up the filtering step and use the Fast Hough Transform to speed up the back projection step. Experimental results on phantom demonstrate computational costs gain with acceptable quality.

\newpage
\bibliographystyle{IEEEbib}
\bibliography{refs}

\end{document}